\documentclass[11pt,a4paper]{article}
\usepackage[margin=25mm]{geometry}
\usepackage{amsmath,amssymb,graphicx,booktabs,array}
\usepackage[numbers,sort&compress]{natbib}
\usepackage[expansion=false]{microtype}
\usepackage{caption}
\usepackage{placeins}
\usepackage[colorlinks=true,linkcolor=blue,citecolor=blue,urlcolor=blue]{hyperref}
\newcommand{\GeV}{\,\mathrm{GeV}}
\newcommand{\TeV}{\,\mathrm{TeV}}
\newcommand{\mchi}{m_\chi}
\newcommand{\om}{\Omega_\chi h^2}

\newcommand{\sigSI}{\sigma^p_{\mathrm{SI}}}

\newcommand{\MassMin}{44}
\newcommand{\MassMax}{2398}

\newcommand{\MedianQ}{0.0522}

\title{Solar Neutrino Signals and Capture--Annihilation Equilibrium of SU(5) Neutralino Dark Matter}
\author{M. Adeel Ajaib$^{1}$ and Abdul Rehman$^{2}$\\[5pt]
\small $^{1}$Pennsylvania State University, Abington College,\\
\small Abington, PA 19001, USA\\[3pt]
\small $^{2}$Department of Physics, University of Delaware,\\
\small Newark, DE 19716, USA}
\date{}
\begin{document}
\maketitle
\begin{abstract}
Motivated by IceCube solar searches, we investigate neutralino dark matter in supersymmetric SU(5) with nonuniversal gaugino masses. Combining micrOMEGAs relic abundances with DarkSUSY calculations, we develop a consistent picture of solar neutrino and muon signals from thermal neutralino subcomponents. We identify how local abundance, neutralino composition, and capture--annihilation equilibrium shape these signals beyond simple scaling with the dark matter fraction. Integrated fluxes and representative benchmarks connect these predictions to published IceCube reference limits. Our results provide a framework for interpreting solar searches and assessing the signals of subdominant neutralino dark matter.
\end{abstract}

\section{Introduction}
The lightest neutralino is a well-motivated dark matter candidate in supersymmetric theories with conserved $R$ parity. Its abundance and detection signatures depend on its bino, wino, and Higgsino components and the surrounding spectrum~\cite{Jungman1996}. Nonuniversal gaugino masses in supersymmetric SU(5) allow these properties to vary substantially~\cite{Ajaib2017SU5}.

IceCube searches for neutrinos from dark matter annihilation in the Sun provide a direct experimental motivation for this study, with its ten-year analysis probing WIMP scattering through solar neutrino observations~\cite{IceCube2025}. Neutralinos can lose energy through scattering in the Sun and become gravitationally bound; their subsequent annihilation produces neutrinos that may reach the detector. Connecting IceCube searches to a neutralino model therefore requires both its capture rate and the evolution toward capture--annihilation equilibrium~\cite{PressSpergel1985,DarkSUSY6,Ellis2010Solar}.

A thermally underabundant neutralino can be treated as a full-density candidate if another cosmological mechanism supplies the missing abundance, or as a subcomponent whose local density follows its thermal fraction. In the latter case, reduced capture lengthens the equilibration time, so the solar signal need not scale linearly with abundance. Relic-fraction rescaling and its connection to solar equilibration have been studied previously~\cite{Duda2002Subcomponent,Duda2003Indirect}. The earlier flipped-SU(5) study of Ajaib, Gogoladze, and Shafi also retained the equilibration timescale in its solar-muon predictions~\cite{Ajaib2011Muon}.

Here we study conventional SU(5) boundary conditions and connect neutralino composition to the departure from linear abundance scaling. We compare full-density and thermal-subcomponent predictions, construct a general abundance--timescale map, and present integrated neutrino and muon signals in the context of IceCube solar searches. The map follows the standard population equation and can be applied independently of the composition weights in this scan.

\section{SU(5) framework and numerical inputs}\label{sec:model}

We adopt the conventional SU(5) matter assignments of the preceding nonuniversal-gaugino framework~\cite{Ajaib2017SU5}. At the unification scale, the scalar soft masses satisfy
\begin{equation}
 m_Q=m_{U^c}=m_{E^c}=m_{10},\qquad
 m_{D^c}=m_L=m_{\overline5}.
\end{equation}
The setup takes common inputs for each representation across the three families. The Higgs and trilinear inputs obey
\begin{equation}
 m_{H_u}=m_{H_d}=m_{\overline5},\qquad
 A_0=a_0m_{\overline5},
\end{equation}
where the Higgs quantities in this equation denote input mass parameters, rather than their weak-scale squared soft masses. The gaugino masses are independent, with
\begin{equation}
 M_1>0,\qquad M_2>0,\qquad M_3<0,\qquad \mu>0.
\end{equation}
The nominal scan domains follow Table 1 of our preceding SU(5) analysis~\cite{AjaibNasir2026SU5}:
\begin{align}
0\leq m_{10},m_{\overline5}&\leq30\TeV, &0\leq M_1,M_2&\leq5\TeV,\notag\\
-5\TeV\leq M_3&\leq0, &-3\leq a_0&\leq3,\notag\\
2\leq\tan\beta&\leq60.&&
\end{align}
The strict signs above apply to the selected sample. ISAJET/ISASUGRA spectra~\cite{ISAJETManual} are passed through SLHA files to micrOMEGAs for the relic abundance and to DarkSUSY for solar rates. We use $m_t=173.3\GeV$, require $\om\leq0.12$, and classify composition at $90\%$ purity.

We write the lightest neutralino as
\begin{equation}
 \widetilde\chi_1^0=N_{11}\widetilde B+N_{12}\widetilde W^0
                +N_{13}\widetilde H_d^0+N_{14}\widetilde H_u^0,
\end{equation}
and define $f_B=|N_{11}|^2$, $f_W=|N_{12}|^2$, and
$f_H=|N_{13}|^2+|N_{14}|^2$. A component is termed bino-like, wino-like, or Higgsino-like when its fraction is at least $0.90$. Otherwise we label the neutralino mixed and identify its two largest components.

The baseline spectrum and flavor selection is
\begin{align}
 123\GeV &\leq m_h\leq127\GeV, &
 2.90\times10^{-4}&\leq\mathrm{BR}(b\to s\gamma)\leq3.87\times10^{-4},\notag\\
 1.95\times10^{-9}&\leq\mathrm{BR}(B_s\to\mu^+\mu^-)\leq3.43\times10^{-9}, &
 0.15&\leq R_{B\tau\nu}\leq2.41,
\end{align}
with $R_{B\tau\nu}=\mathrm{BR}(B_u\to\tau\nu)/\mathrm{BR}(B_u\to\tau\nu)_{\rm SM}$. The baseline mass requirements are $m_{\widetilde g}\geq2.2\TeV$, $m_{\widetilde t_1}\geq700\GeV$, first- and second-generation squark masses of at least $2\TeV$, $m_{\widetilde\chi_1^\pm}\geq103.5\GeV$, and $m_{\widetilde\tau_1}\geq93.2\GeV$. 

We retain a neutralino dark matter candidate with a physical DarkSUSY setup and a successful micrOMEGAs calculation satisfying $0\leq\om\leq0.12$, using the observed abundance as the reference~\cite{Planck2018,micromegasIndirect,micromegasNeutrinos}. The solar selection requires $\mchi\geq20\GeV$, finite physical rates and timescales, successful integrated-rate statuses, and consistency with the population equation below. The selected sample is predominantly wino-like, with smaller bino-like, Higgsino-like, and mixed components.

The particle calculations are used without an additional Sommerfeld correction~\cite{Beneke2016Wino,Beneke2020Wino}. This omission is particularly relevant to the wino-like population: a consistent treatment modifies both the freeze-out abundance, and hence $\xi$, and the low-velocity annihilation coefficient $C_A$ entering solar evolution. The corrections at freeze-out and in the thermalized solar population need not coincide. At fixed $\xi$ and capture rate, an enhanced $C_A$ shortens the equilibration timescale and moves the response toward linear abundance scaling, whereas a reduced relic fraction lowers capture and can oppose this tendency. A definite net shift therefore requires a joint calculation rather than a common multiplicative correction to the flux. The abundance--timescale relation derived below remains valid for consistently recalculated, density-independent coefficients under the stated population assumptions; the numerical sample predictions are conditional on the particle calculations used here.

Solar rates are calculated with DarkSUSY 6.4.3 and its MSSM particle module~\cite{DarkSUSY6}. They use a local density $\rho_0=0.3\GeV\,\mathrm{cm}^{-3}$, the default halo prescription, and a full local neutralino fraction. The solar signal entries specify an energy threshold of $1\GeV$ and an angular cone of $30^\circ$. We compare
\begin{align}
 \rho_\chi^{\rm full}&=\rho_0,\notag\\
 \rho_\chi^{\rm thermal}&=\xi\rho_0,\qquad
 \xi=\frac{\om}{0.12}\leq1.\label{eq:xi}
\end{align}
The second relation assumes that the local and cosmological component fractions agree and that the halo velocity distribution is unchanged. The absolute capture rates also depend on the halo speed distribution through the capture kinematics; we do not vary this distribution, so the numerical predictions are conditional on the adopted halo prescription~\cite{DarkSUSY6}. For an underabundant neutralino, the first scenario requires an additional source of abundance if it is interpreted cosmologically. We do not specify such a source.

The scan distributions and percentages describe this selected sample, rather than probabilities over SU(5) parameter space. The benchmarks illustrate mechanisms under these selections; we do not impose a combined collider and dark-matter likelihood. In particular, the full-density H.E.S.S. selection of the companion study~\cite{AjaibNasir2026SU5} is not applied: subcomponent gamma-ray predictions require the appropriate density-squared normalization, annihilation spectrum, and halo assumptions. Detailed selection and numerical checks are retained in the accompanying analysis.

\section{Capture, annihilation, and abundance rescaling}\label{sec:population}
Let $N$ denote the number of neutralinos gravitationally bound to the Sun. The usual population equation is~\cite{DarkSUSY6,Ellis2010Solar,Ajaib2011Muon}
\begin{equation}
 \frac{dN}{dt}=C_\odot-C_A N^2-C_E N,\qquad
 \Gamma_A=\frac12 C_A N^2,\label{eq:population}
\end{equation}
where $C_\odot$ is the capture rate, $C_A$ is the annihilation coefficient, and $C_E$ describes evaporation. For the retained masses we neglect evaporation. We also assume a thermalized captured population, time-independent coefficients, no self-capture term, and negligible initial population. A self-capture contribution would enter as $+C_sN$, where $C_s$ depends on neutralino self-scattering and the incident halo population. Its importance is governed by $C_sN$ relative to nuclear capture and annihilation, rather than by a large nuclear capture rate alone~\cite{Zentner2009SelfCapture}. We have not calculated $C_s$ for the retained points; the following solution is conditional on self-capture remaining negligible. 

With $C_E=0$ and $N(0)=0$, the solution is
\begin{equation}
 N(t)=\sqrt{\frac{C_\odot}{C_A}}\tanh\left(\frac{t}{\tau}\right),\qquad
 \tau=(C_\odot C_A)^{-1/2},
\end{equation}
At the solar age, this gives
\begin{equation}
 \Gamma_A=\frac12 C_\odot\tanh^2\!\left(\frac{t_\odot}{\tau}\right),\qquad
 \frac{2\Gamma_A}{C_\odot}=\tanh^2\!\left(\frac{t_\odot}{\tau}\right).\label{eq:eqfraction}
\end{equation}
Equilibrium corresponds to $t_\odot/\tau\gg1$ and $\Gamma_A\simeq C_\odot/2$. We use the DarkSUSY evolution time $t_\odot=1.5\times10^{17}\,\mathrm{s}$ and summarize near-equilibrium models by $2\Gamma_A/C_\odot\geq0.95$. The rescaling uses the calculated timescale, avoiding inversion of numerically saturated equilibrium fractions.

At fixed particle parameters and halo velocity distribution, capture is linear in the incident density. Holding $C_A$ fixed therefore gives
\begin{equation}
 C_\odot(\xi)=\xi C_\odot(1),\qquad
 \tau(\xi)=\frac{\tau(1)}{\sqrt{\xi}}.\label{eq:rescale}
\end{equation}
Here and below, the argument 1 denotes full local density and $\xi$ denotes the thermal local fraction. Substitution into Eq.~\eqref{eq:eqfraction} yields
\begin{equation}
 \frac{\Gamma_A(\xi)}{\Gamma_A(1)}
 =\xi\,\frac{\tanh^2[\sqrt{\xi}\,t_\odot/\tau(1)]}
                  {\tanh^2[t_\odot/\tau(1)]}.\label{eq:R}
\end{equation}
For a fixed annihilation spectrum and propagation prescription, the same ratio multiplies an annihilation-induced neutrino or muon flux.

Equation~\eqref{eq:R} has two familiar limits. When both populations are far from equilibrium, expanding the hyperbolic tangent gives $\Gamma_A(\xi)/\Gamma_A(1)\simeq\xi^2$. When both are equilibrated, the ratio approaches $\xi$. Thus,
\begin{equation}
 \xi^2\leq\frac{\Gamma_A(\xi)}{\Gamma_A(1)}\leq\xi\qquad(0<\xi\leq1).\label{eq:bounds}
\end{equation}
A point close to equilibrium at full density can move away from it once capture is reduced. It is therefore insufficient to infer linear solar rescaling from equilibrium at full density alone.

\section{Results}
\subsection{Capture and equilibrium}
\begin{figure}[!t]
 \centering\includegraphics[width=\linewidth]{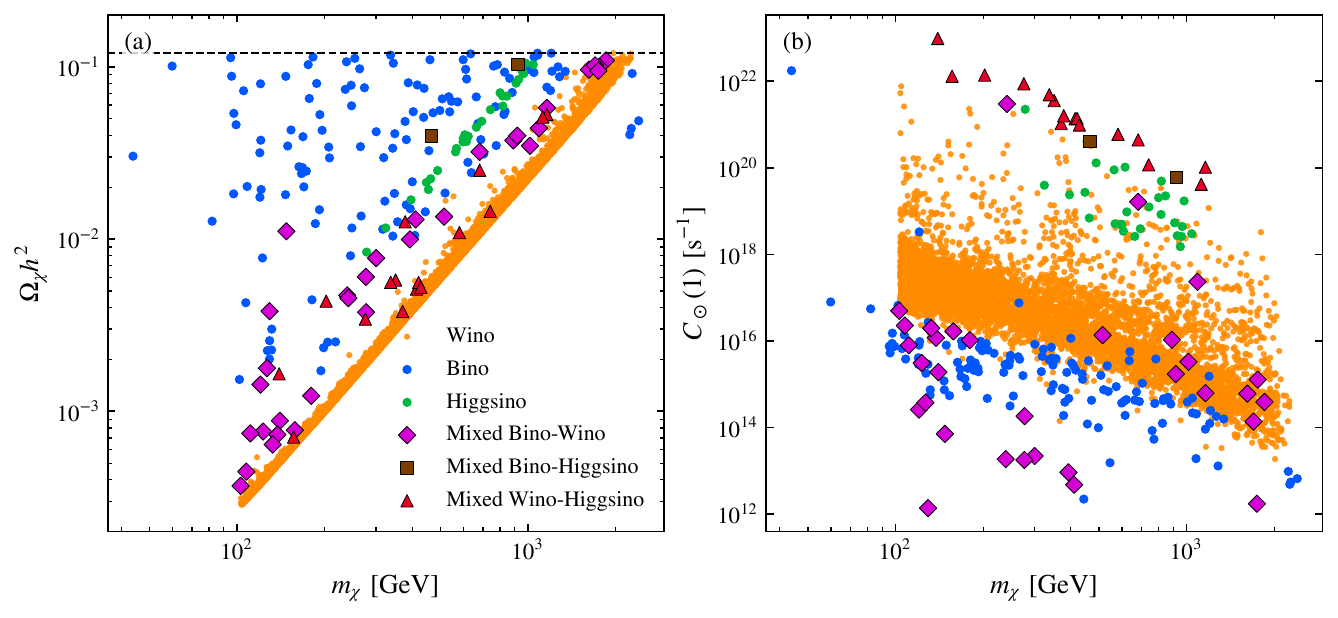}
 \caption{Relic abundance (left) and full-density solar capture rate (right) versus neutralino mass. Colors and markers identify the neutralino composition using the $90\%$ criterion. The dashed line is $\om=0.12$. The same selected sample is used throughout the scatter plots.}\label{fig:sample}
\end{figure}

Figure~\ref{fig:sample} shows relic abundance and full-density capture over the selected mass range $\MassMin$--$\MassMax\GeV$. Most models are thermally underabundant. Capture spans many orders of magnitude even at similar mass, reflecting the composition-dependent scattering amplitudes. Both spin-independent and spin-dependent interactions enter the calculation.

\begin{figure}[!t]
 \centering\includegraphics[width=\linewidth]{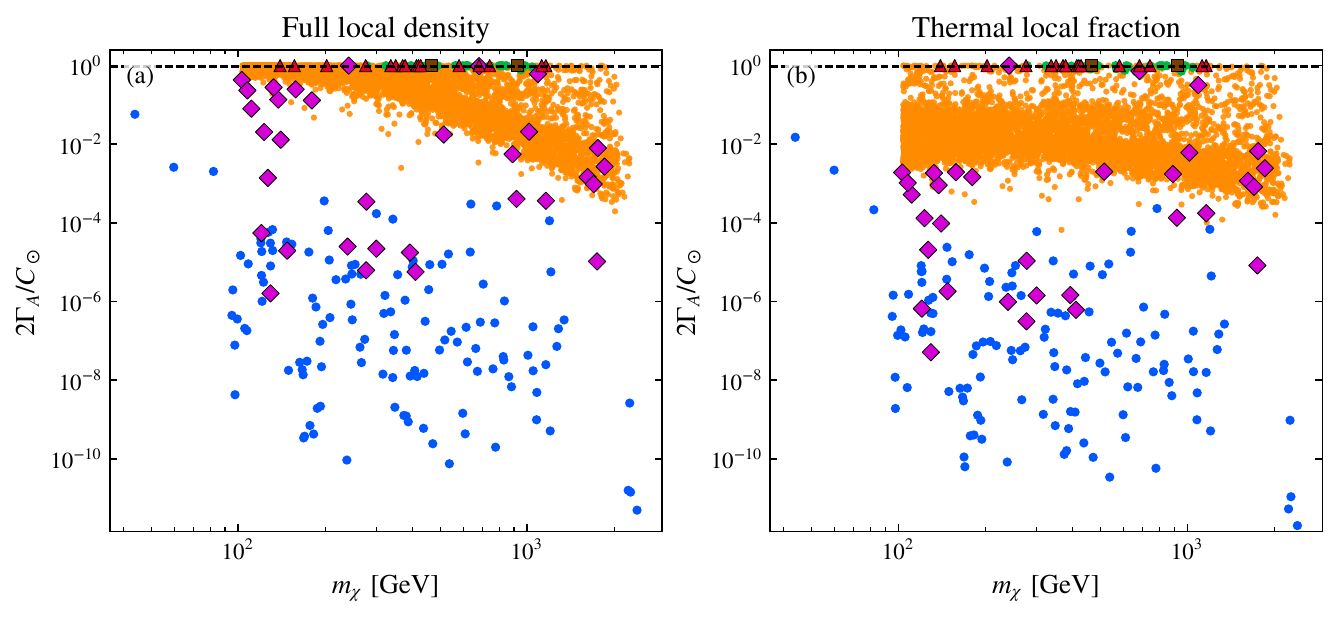}
 \caption{Ratio $2\Gamma_A/C_\odot$ at full local density (left) and at the thermal local fraction (right). The dashed line marks $2\Gamma_A/C_\odot=0.95$. Each point on the right is obtained from the corresponding point on the left using Eqs.~\eqref{eq:eqfraction} and~\eqref{eq:rescale}.}\label{fig:equilibrium}
\end{figure}

Figure~\ref{fig:equilibrium} compares $2\Gamma_A/C_\odot$ at full density and at the thermal fraction. Reducing the abundance shifts models away from equilibrium, although near-equilibrium solutions remain. Table~\ref{tab:composition_response} quantifies this change by composition.

\begin{figure}[!t]
 \centering\includegraphics[width=\linewidth]{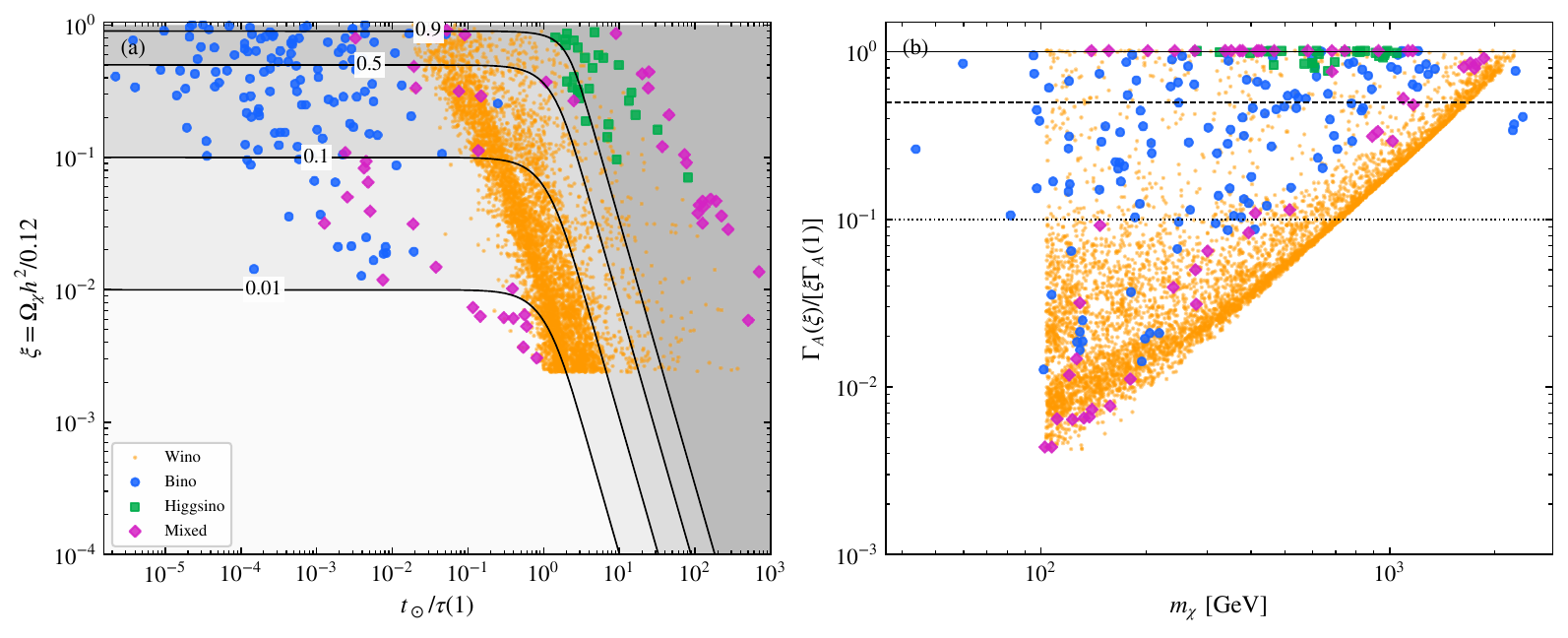}
 \caption{Left: contours of $\Gamma_A(\xi)/[\xi\Gamma_A(1)]$ in the full-density timescale ratio and thermal abundance fraction, with the sample overlaid by composition. Right: this ratio versus mass. The horizontal lines at 1, $1/2$, and $1/10$ denote linear scaling and overestimates by factors of two and ten.}\label{fig:rescaling}
\end{figure}

Figure~\ref{fig:rescaling} separates the abundance dependence from the sample composition. The left panel maps the departure from linear scaling in $t_\odot/\tau(1)$ and $\xi$; its contours follow the population equation independently of the scan. The right panel shows the same ratio for the selected models:

\begin{equation}
 \frac{\Gamma_A(\xi)}{\xi\Gamma_A(1)}
 =\frac{\tanh^2[\sqrt{\xi}\,t_\odot/\tau(1)]}
       {\tanh^2[t_\odot/\tau(1)]}.\label{eq:linear_comparison}
\end{equation}

Unity denotes linear scaling; a ratio below $1/2$ means that this approximation overpredicts the rate by more than a factor of two. This occurs for most selected models, with a median ratio of $\MedianQ$. Models already far from equilibrium approach quadratic abundance scaling, whereas those initially near equilibrium can enter the transition regime as capture decreases.

Table~\ref{tab:composition_response} separates the response by neutralino composition. The predominantly wino-like sample shows strong additional suppression relative to linear scaling. Bino-like states remain far from equilibrium, while the sampled Higgsino-like states lie closer to the linear limit. Composition alone does not fix the correction: the relic fraction and full-density equilibration time jointly determine it through Eq.~\eqref{eq:linear_comparison}. The smaller non-wino populations should be interpreted as illustrative samples.

\begin{table}[!t]\centering\small
\caption{Composition-resolved response before and after the graphical LZ selection. The rate ratio is $\Gamma_A(\xi)/[\xi\Gamma_A(1)]$. Percentages refer to each row; near equilibrium means $2\Gamma_A/C_\odot\geq0.95$. A dash denotes an empty class. Class sizes and cutflow are provided in the accompanying data.}
\label{tab:composition_response}
\setlength{\tabcolsep}{4pt}
\begin{tabular}{lrrrrr}\toprule
Class & Median $\xi$ & Median ratio & Below $1/2$ [\%] & Full near eq. [\%] & Thermal near eq. [\%]\\
\midrule
\multicolumn{6}{l}{\textit{Baseline}}\\
All & 0.0224 & 0.0522 & 90.7 & 19.9 & 1.9\\
Wino & 0.0209 & 0.0498 & 92.0 & 19.8 & 1.4\\
Bino & 0.375 & 0.375 & 61.7 & 0.0 & 0.0\\
Higgsino & 0.354 & 0.975 & 0.0 & 82.1 & 57.1\\
Mixed & 0.0492 & 0.636 & 48.0 & 40.0 & 38.0\\
\midrule
\multicolumn{6}{l}{\textit{LZ reference}}\\
All & 0.0214 & 0.0506 & 92.4 & 18.7 & 0.6\\
Wino & 0.0202 & 0.0489 & 93.1 & 19.2 & 0.6\\
Bino & 0.384 & 0.384 & 61.4 & 0.0 & 0.0\\
Higgsino & -- & -- & -- & -- & --\\
Mixed & 0.0574 & 0.0574 & 80.0 & 0.0 & 0.0\\
\bottomrule\end{tabular}\end{table}

\subsection{Solar signals and IceCube comparisons}\label{sec:flux}\label{sec:theory}\label{sec:icecube}
We calculate integrated solar $\nu_\mu+\bar\nu_\mu$ and through-going $\mu^-+\mu^+$ fluxes above $1\GeV$ within a $30^\circ$ cone. DarkSUSY includes neutrino propagation, interactions, and oscillations, together with muon production and propagation for the muon observable~\cite{WimpSim,DarkSUSY6}. Its method-5 capture prescription integrates the halo velocity distribution and nuclear form factors over solar elements and isotopes.

\begin{figure}[!t]
 \centering\includegraphics[width=\linewidth]{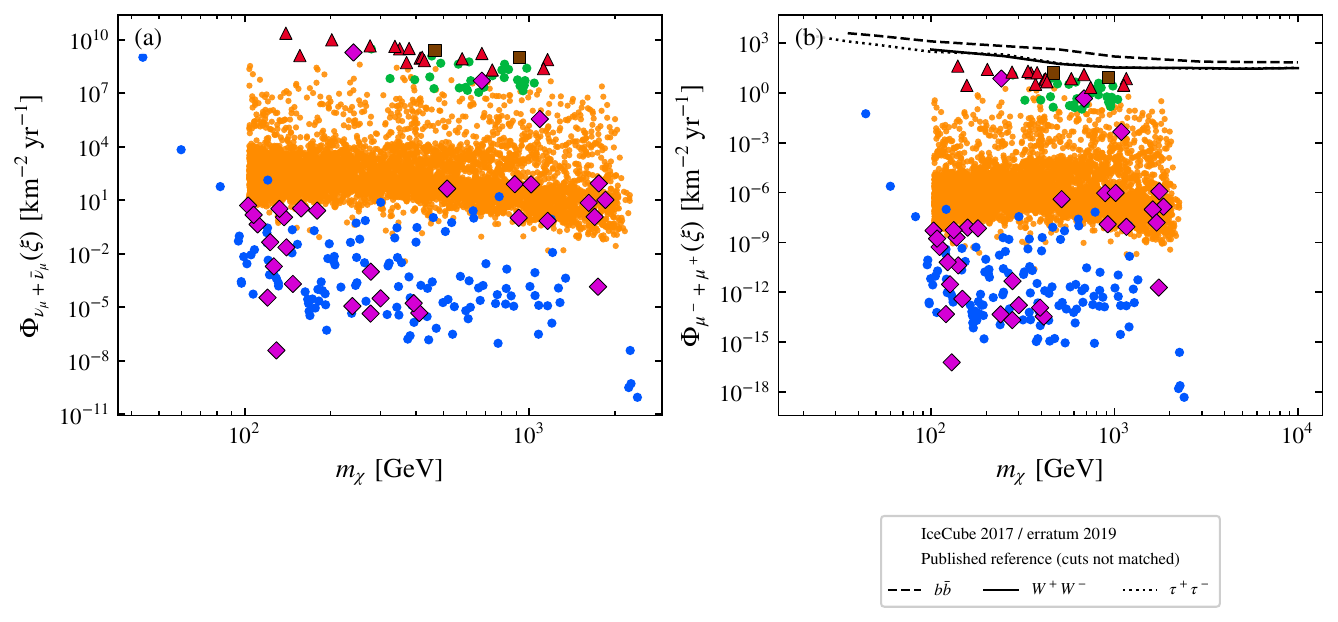}
 \caption{Thermal-fraction integrated $\nu_\mu+\bar\nu_\mu$ flux (left) and through-going $\mu^-+\mu^+$ flux (right), above $1\GeV$ within a $30^\circ$ cone. The right panel includes the corrected IceCube pure-channel muon-flux limits~\cite{IceCube2017Solar,IceCube2019MuonErratum}. The comparison conventions are discussed in Sec.~\ref{sec:icecube}; the panels have separate vertical scales.}\label{fig:fluxes}
\end{figure}

The fluxes and equilibrium calculation use the same capture prescription. We rescale the direct DarkSUSY integrals by Eq.~\eqref{eq:R}, rather than reintegrating sampled differential spectra. Figure~\ref{fig:rescaling} quantifies the suppression relative to linear abundance rescaling, while Fig.~\ref{fig:fluxes} shows the resulting thermal-abundance neutrino and muon fluxes. The muon-flux predictions lie below the displayed IceCube pure-channel reference curves over their tabulated mass ranges. These curves provide experimental context; differences in annihilation spectra and signal selections prevent a model-specific exclusion statement.

\begin{figure}[!t]
 \centering\includegraphics[width=\linewidth]{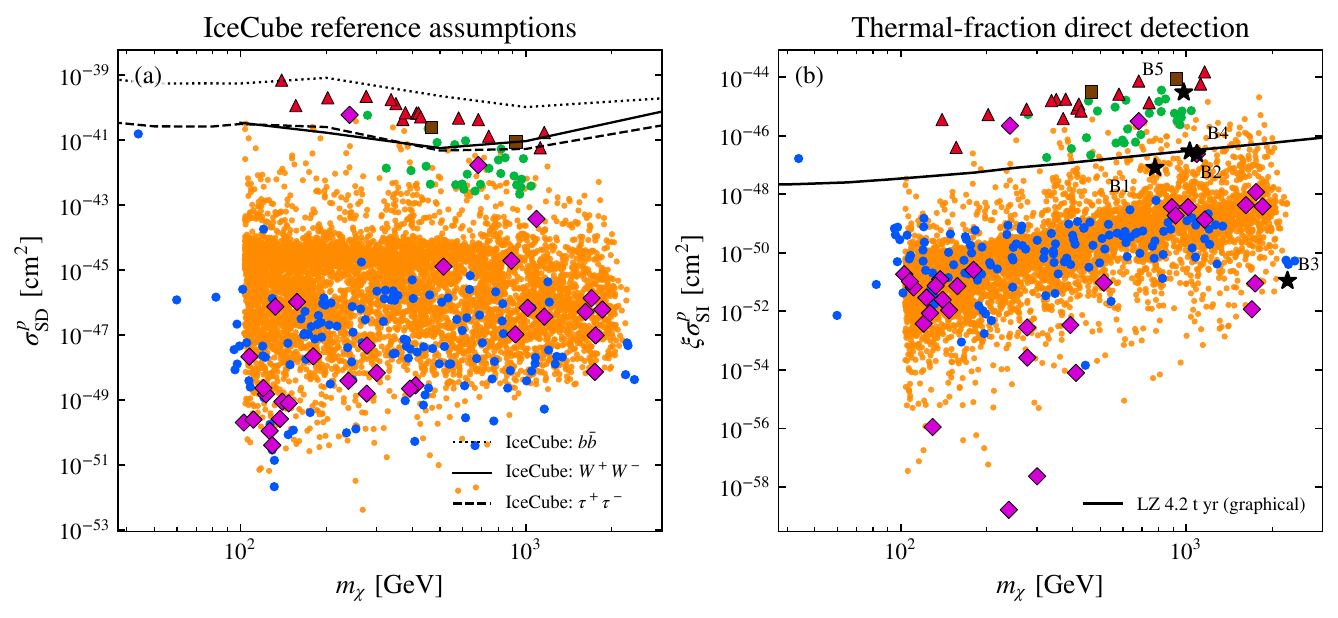}
 \caption{Left: unscaled proton SD cross sections and the 2025 IceCube pure-channel reference limits. Right: $\xi\sigSI$ and the graphical LZ reference, with benchmark stars. The different density conventions reflect the nonlinear solar response and linear direct-detection recoil rate. Comparison assumptions are given in Sec.~\ref{sec:icecube}.}\label{fig:icecube}
\end{figure}

We use the muon-flux limits from the three-year search, corrected in the 2019 erratum~\cite{IceCube2017Solar,IceCube2019MuonErratum}, for Fig.~\ref{fig:fluxes}, and the 2025 ten-year scattering limits~\cite{IceCube2025,IceCube2025Data} for Fig.~\ref{fig:icecube}. The former provide explicit muon-flux limits; the latter supply the scattering-limit curves used here. The overlays retain the published pure-channel conventions, with energy thresholds and angular acceptance not matched to our predictions. A model-specific test requires mixed-channel spectra, annihilation normalization, and detector response~\cite{IceCube79Likelihood}. The conversion from a pure-channel reference to a mixed-channel constraint depends on the neutralino mass, branching fractions, and energy-dependent detector response; no universal numerical conversion factor is applied here.

Figure~\ref{fig:icecube} also compares $\xi\sigSI$ with the graphical LZ reference~\cite{LZ2025Review}, assuming compatible halo conventions and approximately equal proton/neutron SI couplings. Unlike the solar signal, the direct-detection recoil rate scales linearly with local density. We retain models below this curve within its tabulated mass range as an observational cross-check. Table~\ref{tab:composition_response} shows that the departure from linear solar scaling persists in this subset.

\section{Representative benchmarks}\label{sec:benchmarks}
Tables~\ref{tab:model} and~\ref{tab:solar} illustrate the composition dependence of the solar response. B1 is a wino that moves away from near equilibrium after abundance rescaling; B2 is a TeV wino closer to equilibrium but still below the near-equilibrium criterion at its thermal abundance. B3 illustrates weak bino equilibration with heavy sleptons and a compressed neutralino--gluino spectrum. B4 has appreciable capture through its mixed bino--wino composition, while B5 illustrates nearly linear scaling in a Higgsino-like state.

For B3, $m_\chi=2262.1\GeV$ and $m_{\widetilde g}=2272.5\GeV$, giving a mass difference of $10.4\GeV$. This near degeneracy is consistent with a gluino-coannihilation interpretation of its micrOMEGAs relic abundance~\cite{Ellis2015Gluino}. The spectrum supports this possibility, but a channel-resolved freeze-out calculation is needed to establish its contribution.

B1, B3, and B4 lie below the graphical LZ reference. B2 is borderline, with a reference ratio of 0.946; B5 lies above the graphical LZ reference and is retained solely to illustrate nearly linear abundance scaling, without a claim of experimental viability. The small B1 and B2 chargino--neutralino mass differences are the spectrum inputs, not radiatively corrected wino splittings~\cite{Ibe2013Wino}, and are not used to infer lifetimes. Higgs invisible decays are kinematically closed for these benchmarks but are not included in the full-sample selection.

\begin{table}[!t]
\centering
\small
\caption{Benchmark parameters and spectra. Masses are in GeV except the GUT-scale inputs labelled in TeV; $\mu(Q)$ is evaluated at the spectrum renormalization scale. B1 and B2 are wino-like, B3 bino-like, B4 mixed bino--wino, and B5 Higgsino-like. The mass splittings are the spectrum inputs.}
\label{tab:model}
\setlength{\tabcolsep}{5pt}
\begin{tabular}{lrrrrr}
\toprule
Quantity & B1 & B2 & B3 & B4 & B5 \\
\midrule
\multicolumn{6}{l}{\textit{GUT-scale inputs}}\\[2pt]
$m_{10}$ [TeV] & $13.9$ & $12.9$ & $22.2$ & $4.53$ & $7.72$ \\
$m_{\overline5}$ [TeV] & $12.9$ & $12.3$ & $16.3$ & $5.66$ & $9.11$ \\
$M_1$ [TeV] & $3.84$ & $3.7$ & $4.71$ & $2.31$ & $2.33$ \\
$M_2$ [TeV] & $0.838$ & $1.09$ & $3.04$ & $1.15$ & $3.02$ \\
$M_3$ [TeV] & $-0.91$ & $-1.62$ & $-0.815$ & $-4.12$ & $-3.48$ \\
$A_0$ [TeV] & $0$ & $-7.29$ & $-5.31$ & $0.882$ & $8.16$ \\
$\tan\beta$ & $11.8$ & $26.7$ & $16.2$ & $27.5$ & $19.9$ \\
\midrule
\multicolumn{6}{l}{\textit{Weak-scale spectrum and composition}}\\[2pt]
$m_{\widetilde g}$ & $2429$ & $3917$ & $2273$ & $8438$ & $7425$ \\
$m_\chi$ & $777$ & $1028$ & $2262$ & $1089$ & $980$ \\
$m_{\widetilde\chi_1^\pm}$ & $777$ & $1029$ & $2745$ & $1090$ & $988$ \\
$\Delta m_{\chi^\pm-\chi^0}$ & $0.00122$ & $0.00452$ & $483$ & $0.954$ & $8.36$ \\
$\mu(Q)$ & $2328$ & $2112$ & $10311$ & $2849$ & $964$ \\
$m_h$ & $124$ & $124$ & $125$ & $124$ & $125$ \\
$f_B$ & $6.50\times10^{-7}$ & $4.60\times10^{-6}$ & $1$ & $0.339$ & $0.058$ \\
$f_W$ & $0.999$ & $0.998$ & $5.57\times10^{-6}$ & $0.659$ & $0.00102$ \\
$f_H$ & $0.00144$ & $0.00249$ & $1.50\times10^{-5}$ & $0.00131$ & $0.941$ \\
$\Omega_\chi h^2$ & $0.0138$ & $0.0239$ & $0.0439$ & $0.0441$ & $0.105$ \\
$m_{\widetilde\tau_1}$ & $12843$ & $11898$ & $16158$ & $4178$ & $7345$ \\
$m_{\widetilde e_L}$ & $12927$ & $12341$ & $16399$ & $5719$ & $9308$ \\
$m_{\widetilde e_R}$ & $13957$ & $12940$ & $22224$ & $4607$ & $7754$ \\
\bottomrule
\end{tabular}
\end{table}

\begin{table}[!t]\centering\small
\caption{Selected solar observables for the benchmarks in Table~\ref{tab:model}. Capture, annihilation and fluxes use the thermal local abundance; the equilibrium ratio is also shown at full density. Fluxes are above $1\GeV$ within a $30^\circ$ cone. The final rows compare the annihilation rate with its linear abundance estimate and the rescaled SI cross section with the graphical LZ reference.}
\label{tab:solar}
\setlength{\tabcolsep}{4pt}
\begin{tabular}{lrrrrr}\toprule
Quantity & B1 & B2 & B3 & B4 & B5 \\
\midrule
$C_\odot$ [s$^{-1}$] & $5.63\times10^{16}$ & $1.18\times10^{17}$ & $1.78\times10^{12}$ & $8.55\times10^{16}$ & $1.50\times10^{19}$ \\
$2\Gamma_A/C_\odot$ (full density) & $0.975$ & $0.981$ & $2.62\times10^{-9}$ & $0.615$ & $1$ \\
$2\Gamma_A/C_\odot$ (thermal) & $0.485$ & $0.69$ & $9.57\times10^{-10}$ & $0.32$ & $0.999$ \\
$\Gamma_A$ [s$^{-1}$] & $1.37\times10^{16}$ & $4.06\times10^{16}$ & $854$ & $1.37\times10^{16}$ & $7.52\times10^{18}$ \\
$\Phi_{\nu_\mu+\bar\nu_\mu}$ [km$^{-2}$ yr$^{-1}$] & $3.69\times10^{5}$ & $1.04\times10^{6}$ & $3.80\times10^{-8}$ & $3.59\times10^{5}$ & $2.53\times10^{8}$ \\
$\Phi_{\mu^-+\mu^+}$ [km$^{-2}$ yr$^{-1}$] & $0.00413$ & $0.0128$ & $2.43\times10^{-16}$ & $0.00448$ & $2.57$ \\
Rate / linear estimate & $0.498$ & $0.703$ & $0.366$ & $0.519$ & $1$ \\
LZ reference ratio & $0.342$ & $0.946$ & $1.72\times10^{-5}$ & $0.732$ & $103$ \\
\bottomrule\end{tabular}\end{table}

\FloatBarrier

\section{Conclusions}
A reduced neutralino abundance lowers solar capture and lengthens the equilibration timescale. Consequently, the annihilation rate interpolates between linear and quadratic abundance scaling. In the selected SU(5) sample, linear rescaling substantially overestimates the solar signal for many models. The response depends jointly on composition, abundance, and equilibration time, and persists after the graphical direct-detection reference selection.

The integrated neutrino and muon predictions connect this response to IceCube solar searches, with published limits providing experimental reference comparisons. The general abundance--timescale map also applies to other consistently calculated particle inputs. Interpreting IceCube searches for subdominant neutralinos requires evolving the captured population at the assumed local abundance, since equilibrium at full density does not guarantee linear signal rescaling.

\begingroup
\footnotesize
\setlength{\baselineskip}{9.4pt}
\setlength{\bibsep}{0pt}
\setlength{\parskip}{0pt}
\interlinepenalty=10000
\bibliographystyle{unsrtnat}
\bibliography{references}
\endgroup
\end{document}